\documentclass{emulateapj}

\usepackage{amsmath}
\usepackage{amssymb}
\usepackage{graphicx}
\usepackage[usenames,dvipsnames]{color}
\usepackage{natbib}
\usepackage[backref,breaklinks,colorlinks,citecolor=blue]{hyperref}
\usepackage[all]{hypcap}

\newcommand{\WMAP}{\textsl{WMAP}}

\newcommand{\wmap}{{\WMAP}}
\newcommand{\Planck}{{\textsl{Planck}}}
\newcommand{\planck}{{\textsl{Planck}}}
\newcommand{\lcdm}{\ensuremath{\Lambda}CDM}
\newcommand{\kmsmpc}{\ensuremath{{\rm km\,s}^{-1}{\rm Mpc}^{-1}}}
\newcommand{\rsdv}{\ensuremath{r_d/D_V}}

\newcommand{\dvrfidrd}{\ensuremath{D_V\, r_{d,\rm fid}/r_d}}

\renewcommand{\ell}{\ensuremath{l}}
\newcommand{\be}{\begin{equation}}
\newcommand{\ee}{\end{equation}}

\newcommand{\Hubble}{\ensuremath{69.6 \pm 0.7} \kmsmpc}
\newcommand{\HubbleMean}{\ensuremath{69.6}}
\newcommand{\HubbleSigma}{\ensuremath{0.7}}

\def\ba{\begin{eqnarray}}
\def\ea{\end{eqnarray}}

\newcommand{\barr}{\begin{array}}
\newcommand{\earr}{\end{array}}

\renewcommand{\S}{Section~}

\begin{document}

\title{The 1\% Concordance Hubble Constant}

\author{C. L. Bennett, D. Larson, and J. L. Weiland}
\affil{Johns Hopkins University}
\affil{3400 N. Charles St, Baltimore, MD 21218}
\email{cbennett@jhu.edu}
\and
\author{G. Hinshaw}
\affil{Dept. of Physics and Astronomy, University of British Columbia}
\affil{Vancouver, BC  Canada V6T 1Z1}

\begin{abstract}

The determination of the Hubble constant has been a central goal in observational astrophysics 
for nearly 100 years. Extraordinary progress has occurred in recent years on two fronts: the 
cosmic distance ladder measurements at low redshift and cosmic microwave background (CMB) 
measurements at high redshift. The CMB is used to predict the current expansion rate through a 
best-fit cosmological model. Complementary progress has been made with baryon acoustic 
oscillation (BAO) measurements at relatively low redshifts.  While BAO data do not 
independently determine a Hubble constant, they are important for constraints on possible 
solutions and checks on cosmic consistency. A precise determination of the Hubble constant is 
of great value, but it is more important to compare the high and low redshift measurements to 
test our cosmological model. Significant tension would suggest either uncertainties not 
accounted for in the experimental estimates, or the discovery of new physics beyond the 
standard model of cosmology.  In this paper we examine in detail the tension between the CMB, 
BAO, and cosmic distance ladder data sets. We find that these measurements are consistent 
within reasonable statistical expectations, and we combine them to determine a best-fit 
Hubble constant of \Hubble.  
This value is based upon \wmap9+SPT+ACT+6dFGS+BOSS/DR11+$H_0$/Riess; we explore 
alternate data combinations in the text.
The combined data constrain the Hubble constant to 1\%, with no 
compelling evidence for new physics.

\end{abstract}

\newpage

\section{Introduction}

Hubble's original attempt to establish the distance-redshift relation \citep{hubble:1929} highlighted the difficulty of establishing accurate distances over a sufficient volume of the universe while controlling systematic measurement errors. This lesson was reinforced in the decades-old debate between de Vaucouleurs and Sandage, who claimed Hubble constant values of $\sim$100 and $\sim$50 \kmsmpc, respectively. In part, this debate motivated the {\it Hubble Space Telescope} ({\it HST}) and its Key Project to determine the Hubble constant, which produced $H_0=72\pm 8$ \kmsmpc\ \citep{freedman/etal:2001}.   Efforts over the past dozen years have produced similar values, but the uncertainties have been reduced from $\sim$11\% to $\sim$3\%: $73.8 \pm 2.4$ \kmsmpc\ \citep{riess/etal:2011}; see also \citet{freedman/madore:2010,freedman/madore:2013} and \citet{livio/riess:2013} for more detailed reviews.  Perhaps not surprisingly, the current measurements of $H_0$ lie near the mean of the de Vaucouleurs and Sandage values. 

In recent years, Hubble constant measurements based on the cosmic distance ladder have been complemented by measurements using the cosmic microwave background (CMB).  By determining the energy budget of the universe from the CMB power spectra (including polarization), the Friedmann equation determines the dynamics of the universe and hence the value of the Hubble constant through the cosmological model.  The nine-year {\it Wilkinson Microwave Anisotropy Probe} (\WMAP) data, taken alone, but assuming a flat universe, give a 3\% determination of $H_0=70.0 \pm 2.2$ \kmsmpc \citep{bennett/etal:2013, hinshaw/etal:2013}.  When \WMAP\ data are combined with smaller angular scale CMB data from the South Pole Telescope (SPT) and the Atacama Cosmology Telescope (ACT), and with distance measurements derived from baryon acoustic oscillation (BAO) observations in the range $0.1 < z < 0.6$, one obtains $H_0=68.76\pm 0.84$ \kmsmpc, a 1.2\% determination.  The \planck\ Collaboration \citep{planck/16:2013} found a six-parameter flat $\Lambda$CDM model to be a good fit to their data and derived best-fit Hubble constant of $67.3 \pm 1.2$ \kmsmpc.  They note a $\approx 2.5\sigma$ tension with the \citet{riess/etal:2011} cosmic distance ladder measurement.  

Since the two approaches to measuring $H_0$ sample different epochs in cosmic history, this apparent tension has led to guarded speculation that new physics beyond the standard cosmological model may be needed to reconcile them \citep[e.g.,][]{wyman/etal:2014,hamann/hasenkamp:2013,battye/moss:2014,dvorkin/etal:2014}. 
However, as noted by others  \citep[e.g.,][]{verde/protopapas/jimenez:2014,gao/gong:2013,efstathiou:2014},
the tension could simply be the result of uncharacterized systematics in the individual $H_0$ determinations, or a statistical fluke, and these possibilities should be explored.  In this paper, we re-examine the consistency of the $H_0$ measurements.  In Section~\ref{sec:data_sources} we describe the recent relevant data; we present our analysis in Section~\ref{sec:analysis}; and we offer conclusions in Section~\ref{sec:conclusions}. 

\newpage 
\section{Data} \label{sec:data_sources}

We consider three primary data sets for establishing constraints on the Hubble constant: cosmic distance scale measurements, BAO measurements, and CMB measurements, and we describe each in turn.

\subsection{Cosmic Distance Scale Measurements}

\citet{riess/etal:2011} found $H_0=73.8 \pm 2.4$ \kmsmpc\ including systematic  uncertainties.  Their result was based on three methods for finding distances to Type Ia supernovae and therefore calibrating their luminosities: (1) a geometric distance to NGC 4258 based on megamaser measurements, (2) parallax distances to 13 Milky Way Cepheids from {\it HST} and {\it Hipparcos} data, and (3) eclipsing binary distances to 92 colocated Cepheids in the Large Magellanic Cloud.   Independently, \cite{freedman/etal:2012} recalibrated the {\it HST} Key Project distance ladder using mid-infrared observations of Cepheids to minimize some of the systematic effects present in optical observations; they obtain $H_0 = 74.3 \pm 1.5$ (statistical) ${\pm 2.1}$ (systematic).   Both \wmap\ \citep{bennett/etal:2013, hinshaw/etal:2013} and \Planck\ \citep{planck/16:2013} adopted the \citet{riess/etal:2011} value for their 2013 analyses that included an $H_0$ prior.

Since the \citet{riess/etal:2011} publication, there have been proposed revisions to the underlying distance calibration and/or to the selection of calibrators.  \citet{humphreys/etal:2013} recalibrated the geometric distance to NGC 4258 and used that distance to calibrate the Cepheids in NGC 4258.  That work resulted in a value of $H_0=72.0\pm 3.0$ \kmsmpc using only NGC 4258 as an anchor.  \citet{efstathiou:2014} makes a different Cepheid selection and uses the \citet{humphreys/etal:2013} distance to NGC 4258 to obtain $H_0 = 70.6\pm3.3$ \kmsmpc, using the NGC 4258 recalibration alone, and $H_0 = 72.5 \pm 2.5$ \kmsmpc\ using all three calibration methods.  \citet{riess:2014} combines the Humphreys recalibration with the other two calibration methods used in the original \citet{riess/etal:2011} to obtain $H_0=73.0\pm 2.4$ \kmsmpc, making use of all available calibrator data and the original \citet{riess/etal:2011} outlier cut methodology.  For our analysis we adopt $H_0$ from \citet{riess:2014}.

\subsection{Baryon Acoustic Oscillation Data}

The remnants of BAOs in the early universe imprint a characteristic scale in the two-point correlation function of large scale structure.  The BAO scale, $r_s$, is the distance a sound wave traveled in the baryon-photon plasma until decoupling at redshift $z_d$; it is common to define $r_d \equiv r_s(z_d)$.   Since the fluid is relativistic, the sound speed is well determined, so the BAO scale depends primarily on the age of the universe at decoupling, which depends only weakly on the cosmic matter and radiation densities \citep{Reid/etal:2002}.

BAO measurements do not, by themselves, constrain $H_0$, but they are valuable for measuring relative distances as a function of redshift.  The angular extent of the BAO scale at a given redshift determines the relative angular diameter distance, $D_A(z)/r_d$, while the radial extent, $\Delta z$, determines the Hubble parameter $H(z)$.  It is convenient to define a volume-averaged effective distance, $D_V \equiv (D_A^2\; cz/H)^{1/3}$ \citep{eisenstein/etal:2005}, and a fiducial value of the BAO scale, $r_{d,\rm fid}$, for a fiducial cosmological model.  One can then quote observational results as a distance, $D_V(z) \cdot(r_{d,\rm fid} / r_d$).  The fiducial distance depends on cosmological parameters (such as $H_0$), however it is only used as a reference value and does not propagate into the BAO results that we analyze.

\capstartfalse
\begin{deluxetable*}{ccccc}
\tablecaption{BAO Measurements\label{tab:bao_data_tab}}
\tabletypesize{\scriptsize}
\tablewidth{0pt}
\tablehead{\colhead{Quantity} & \colhead{$z$} & \colhead{Value} &
\colhead{Experiment} & \colhead{Reference}}
\startdata
$D_V \frac{r_{d,\rm fid}}{r_d}$ & 0.106 & $457\pm 20$ Mpc & 6dFGS & \citet{beutler/etal:2011}\tablenotemark{a} \\
$D_V \frac{r_{d,\rm fid}}{r_d}$ & 0.32 & $1264\pm 25$ Mpc & BOSS DR11 & \citet{anderson/etal:2013}\tablenotemark{b} \\
$D_V \frac{r_{d,\rm fid}}{r_d}$ & 0.57 & $2056\pm 20$ Mpc & BOSS DR11 & \citet{anderson/etal:2013} \\
$D_A/r_d$ & 2.34 & $10.93\pm 0.34$ & BOSS DR11 & \citet{delubac/etal:2014}\tablenotemark{c}\\
$\frac{c}{H r_d}$ & 2.34 & $9.15\pm 0.20$ & BOSS DR11 & \citet{delubac/etal:2014}\tablenotemark{c}\\
\enddata
\tablenotetext{a}{\citet{beutler/etal:2011} quote their result in the form $\rsdv = 0.336\pm0.015$.  This formulation assumes $r_{d}$ follows the fitting formula of \citet{eisenstein/hu:1998} rather than the CAMB version discussed by \citet{anderson/etal:2013}.  We quote their result in this form for consistency with the BOSS DR11 results.} 
\tablenotetext{b}{\citet{anderson/etal:2013} quote the $z=0.32$ results from \citet{tojeiro/etal:2014}.}
\tablenotetext{c}{The Ly-${\alpha}$ forest results do not optimally combine to a value for $D_V$, so we quote separate radial and tangential components, derived from a combination of auto- and cross-correlation statistics.  As described in the text, we use an optimal combination of these values for our $z=2.34$ likelihood.}
\end{deluxetable*}
\capstarttrue

Table~\ref{tab:bao_data_tab} summarizes the most recent BAO data for redshifts of 0.106, 0.32, 0.57 and 2.34.  With the exception of the lowest redshift, these results are from the 11th Baryon Oscillation Sky Survey (BOSS) CMASS (``constant mass'') data release, DR11. There are several papers that analyze these data, and as noted in \citet{beutler/etal:2013}, they produce results within $1\sigma$ of each another.  For the redshift 0.32 and 0.57 results, we adopt the ``consensus'' values from  \citet{anderson/etal:2013}.  

Currently, BAO results at higher redshift ($2.1 \le z \le 3.5$) come from the analysis of quasar Ly-$\alpha$ forest lines \citep{delubac/etal:2014,font-ribera/etal:2013},
and are derived somewhat differently.  Results are reported as separate radial and tangential determinations in the form $\alpha_\parallel \propto c/(H(z) r_{d})$ and $\alpha_\perp \propto D_{A}(z)/r_{d}$, respectively.  The radial measurement is substantially more precise than the tangential one.  
We ``optimally combine'' the radial and tangential Ly-$\alpha$ data as $\alpha_\parallel^{0.7}\alpha_\perp^{0.3}$ \citep{delubac/etal:2014}.
We obtain the value $\alpha_\parallel^{0.7}\alpha_\perp^{0.3} = 1.017 \pm 0.015$ from the combined contours of the cross- and auto-correlation likelihoods in Figure~13 of that paper,
and we use this as our $z=2.34$ likelihood.

\subsection{Cosmic Microwave Background Data}

\subsubsection{\WMAP}

The \WMAP9 data, when used without other measurements, gives a Hubble constant of $H_0 = 70.0\pm 2.2$ \kmsmpc\ for a flat, six-parameter $\Lambda$CDM universe \citep{bennett/etal:2013, hinshaw/etal:2013}.  Adding the ACT, SPT, and BAO data that were available as of mid-December 2012, the value drops somewhat to $H_0 = 68.76 \pm 0.84$ \kmsmpc\ \citep{hinshaw/etal:2013}, independent of the cosmic distance scale measurements.  Adding the \citet{riess/etal:2011} $H_0$ prior of $73.8 \pm 2.4$ \kmsmpc\ to the \WMAP9+ACT+SPT+BAO analysis gives $H_0 = 69.32 \pm 0.80$ \kmsmpc\ \citep{bennett/etal:2013, hinshaw/etal:2013}.  We discuss combinations of \WMAP9 with more recent data sets in \S\ref{sec:analysis}.

\subsubsection{\Planck}
\label{sec:planck}

The \planck\ Collaboration \citep{planck/16:2013} gives a best-fit value of $67.3 \pm 1.2$ \kmsmpc, derived from a six-parameter flat, $\Lambda$CDM model.  Their analysis makes slightly different assumptions from those used by the \wmap\ team.  The \planck\ collaboration assumed a total neutrino mass of $\sum m_\nu = 0.06$ eV, corresponding to a physical neutrino density of $\Omega_\nu h^2 \approx 6 \times 10^{-4}$
\citep{planck/16:2013}.  In addition, they vary the helium fraction, $Y_{\rm He}$, as a function 
of both the physical baryon density, $\Omega_b h^2$, and the number of effective relativistic degrees of freedom in the early universe, $N_{\rm eff}$, per the big bang nucleosynthesis (BBN) prediction, obtained by interpolating the {\tt PArthENoPE} BBN code \citep{planck/16:2013, pisanti/etal:2008}.  
Note that the $Y_{\rm He}$ fluctuations in a Markov chain have a standard deviation less than 0.1\% of the mean 
$Y_{\rm He}$ value, for a simple $\Lambda$CDM model where $N_{\rm eff}=3.046$ is held constant.
Allowing $Y_{\rm He}$ to vary in this fashion does not provide an extra degree of freedom, because $Y_{\rm He}$ is effectively constrained
by $\Omega_b h^2$.
If these BBN assumptions are applied to the \WMAP9 data, the \WMAP9-only $\Lambda$CDM value of $H_0$ is reduced by 0.3 to $69.7 \pm 2.2$ \kmsmpc (based on the \WMAP9 chains released by the \planck\ collaboration).

The results noted above are from the \planck\ team's first -- and at this time only -- cosmological data release of March 2013.  In version 2 of \citet{planck/16:2013}, dated December 2013, it is shown in Appendix~C.4 that noise from a 4~K cooler line produces a systematic dip in the 217 GHz power spectrum around multipole $\ell=1800$.  When they marginalize over a spectrum template intended to mimic this cooler line, the value of the Hubble constant derived from the full data set increases by 0.3$\sigma$. There is no plan to formally update the publicly available Markov chains to include this effect until the next data release, and the inputs required to do so are not currently available (see also \citet{spergel/flauger/hlozek:2013}), thus all \planck-based numbers in this paper are derived from the initially released data.  Given this somewhat uncertain situation, our final estimate of $H_0$ uses \WMAP9+ACT+SPT CMB data, which produces a conservative uncertainty.

\subsubsection{ACT and SPT}
\label{sec:sptact}

Both \WMAP\ and \Planck\ use supplementary data at small angular scales from the two ground-based experiments, ACT and SPT.  For \planck-based results, we use the publicly available \Planck\ chains which incorporate ACT and SPT data available circa 2013.  For \wmap-based results, we briefly quote results from the \wmap9 release \citep{bennett/etal:2013, hinshaw/etal:2013}, which incorporated ACT and SPT data available circa 2012; the new \wmap9-based results presented here incorporate the most recent ACT and SPT likelihoods. 

The current ACT temperature and lensing likelihoods are discussed in \citet{das/etal:2013} and \citet{2013JCAP...07..025D}, while the most recent SPT likelihoods are discussed in \citet{2013ApJ...779...86S} and \citet{2012ApJ...756..142V}.  A code that incorporates both ACT and SPT temperature likelihoods, called {\tt actlite}, is described in Section~5 of \citet{2013JCAP...07..025D}, and has been made available on LAMBDA\footnote{\href{http://lambda.gsfc.nasa.gov/product/act/act_cmblikelihood_get.cfm}{{\tt http://lambda.gsfc.nasa.gov/product/act/act\_cmblikelih}\\ {\tt ood\_get.cfm}}, version 2.2}.  We use this code in the mode where it uses the ACT--E(quatorial) and SPT data, ignoring the ACT-S(outhern) field to avoid double-counting the overlapping SPT and ACT-S fields.  We supplement the {\tt actlite} likelihood with the lensing likelihoods described in \citet{das/etal:2013} and \citet{2012ApJ...756..142V}.

\section{Analysis} \label{sec:analysis}

Our analysis proceeds in two steps:  first, we form various weighted averages of the CMB, BAO, and distance ladder 
data sets to determine the overall best value of $H_0$; these results are summarized in Table~\ref{tab:combination}.
All CMB+BAO results in this table assume a flat, six-parameter $\Lambda$CDM cosmology. 
Then, we test whether the individual measurements are consistent with one another across techniques; we find that they are.

\begin{figure} \begin{center} 
\begin{tabular}{c}
\includegraphics[height=2.5in]{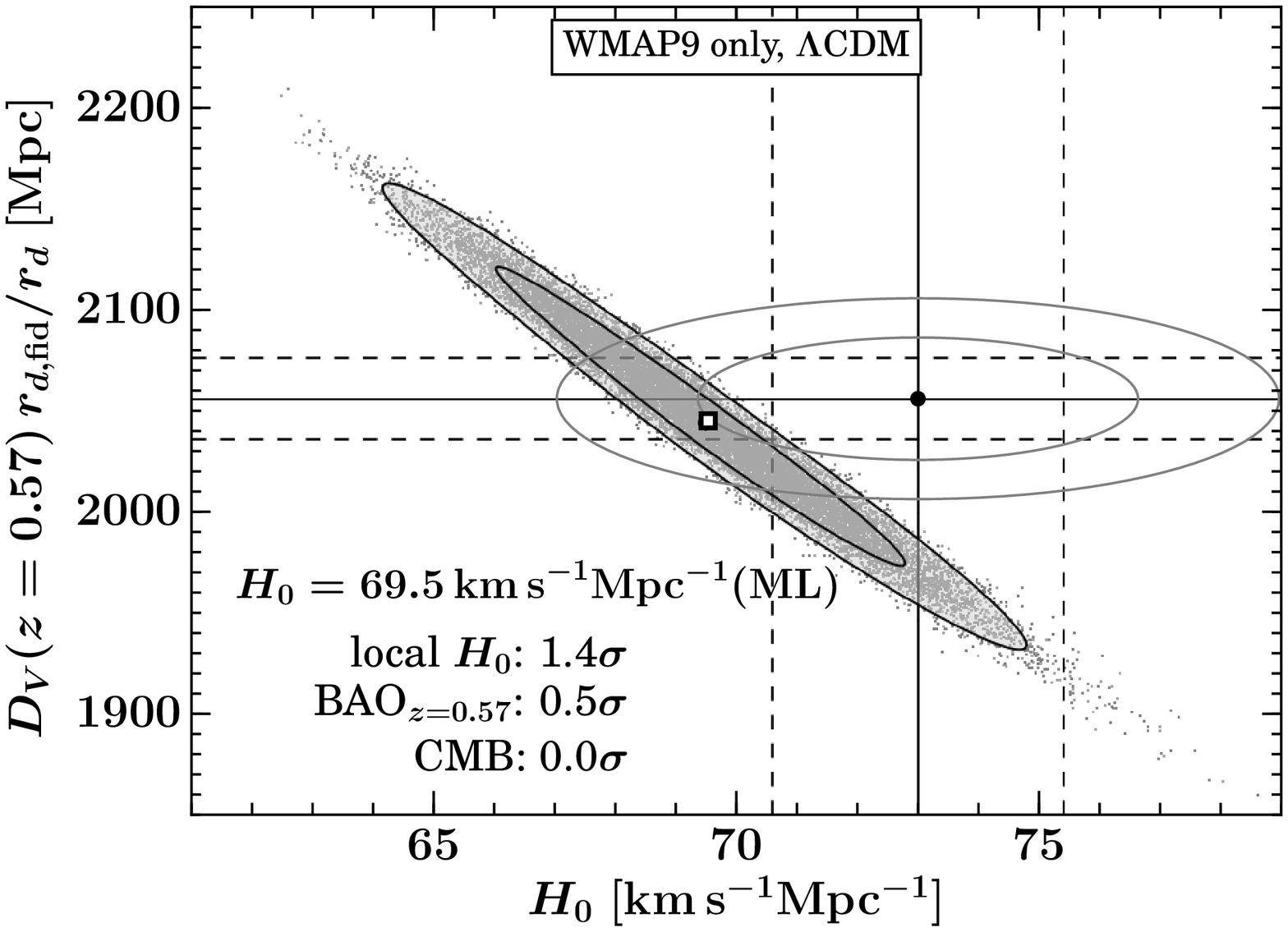} \\[-0.1in]
\includegraphics[height=2.5in]{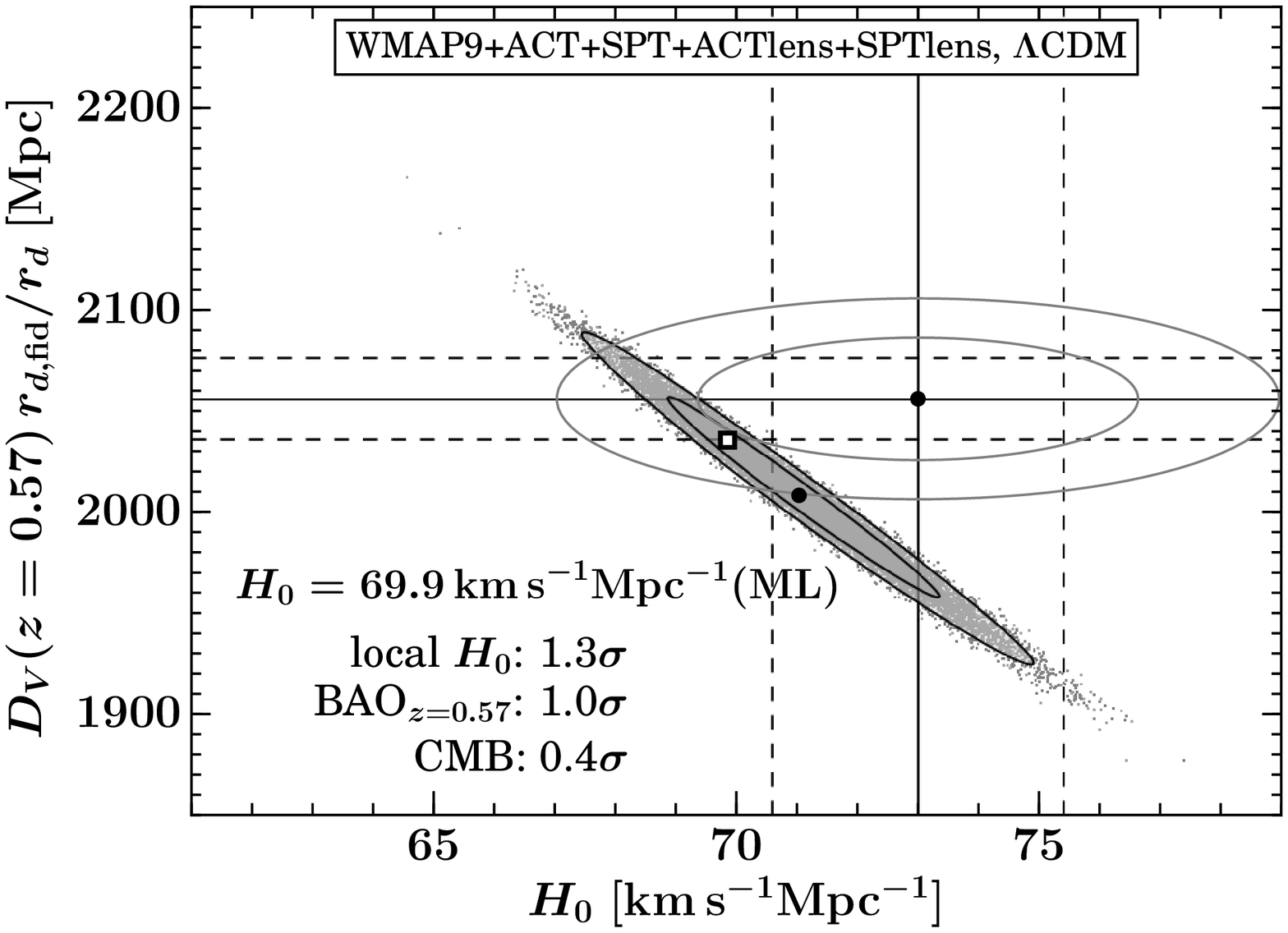} \\[-0.1in]
\includegraphics[height=2.5in]{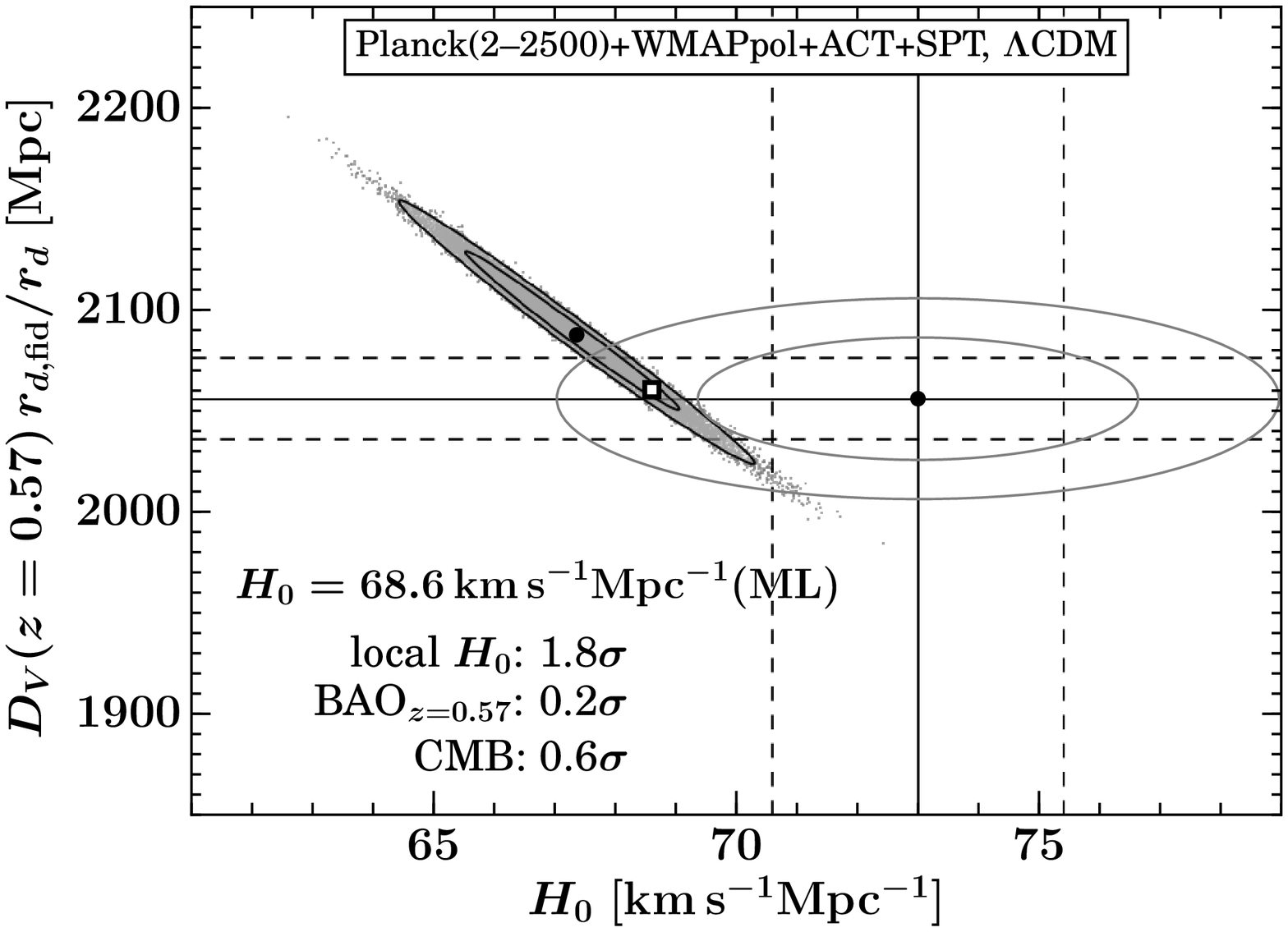} \\[-0.2in]
\end{tabular}
\end{center} 
\caption{
\WMAP9-based (top, middle) and \Planck-based (bottom) CMB data sets for a 
flat $\Lambda$CDM cosmological model, compared to local $H_0$ measurements 
and BAO data at $z=0.57$.  Gray points in the top panel are from this paper's 
\WMAP9/2014 chain; those in the middle are from (\WMAP9+ACT+SPT)/2014.  Gray points 
in the bottom panel are from the \Planck\ chains \citep{planck/16:2013} which 
include the (ACT+SPT)/2013 data combination (see text and Table~2).  
The peak of each 2D-marginalized CMB likelihood is indicated with a black 
filled circle and the contours for the chain points enclose 68\% and 95\% of  
the probability, as computed from a 10-component 2D Gaussian mixture fit 
(see text).  The BAO and $H_0$ measurements are indicated by vertical and 
horizontal lines, respectively, with dashed lines showing $1\sigma$ error bars.
The gray ellipses show 68\% and 95\%
contours for the product of the BAO and $H_0$ likelihoods.  The peak of the 
combined CMB+BAO+$H_0$ likelihoods is indicated with an open square, and its 
value is noted in each panel.  
\label{fig:wmap_planck_lcdm}}
\end{figure}

As noted in \S\ref{sec:planck}, the released \WMAP9/2013 and \Planck\ chains have different assumptions about massive neutrinos and helium.   
For the \WMAP9-based results in Figures~\ref{fig:wmap_planck_lcdm}, \ref{fig:bao_h0_cmb1}, and \ref{fig:bao_h0_cmb2}, we have generated new Markov chains (denoted \wmap9/2014 in Table~\ref{tab:combination}) with $\sum m_\nu = 0.06$ eV and the helium consistency relationship in Equation~8 of \citet{sievers/etal:2013}.  Where noted, we include the most recent ACT and SPT data (\S\ref{sec:sptact}).  We adopt the March 2013 version of the Code for Anisotropies in the Microwave Background (CAMB)\footnote{\url{http://camb.info}}, which is the same version used by \Planck.

\capstartfalse
\begin{deluxetable*}{l|crrrrrrrr}
\tablecaption{Estimates of the Hubble Constant from Combined Data, Assuming $\Lambda$CDM\label{tab:combination}}
\tablewidth{0pt}
\tablehead{\colhead{Data set} & \colhead{$H_0$} & \multicolumn{8}{c}{Combined results} \\ 
& \kmsmpc & 
\colhead{A} & 
\colhead{B} & 
\colhead{C} & 
\colhead{D} & 
\colhead{E} &
\colhead{F} &
\colhead{G} & 
\colhead{H\tablenotemark{a}} }
\tabletypesize{\footnotesize}
\startdata
\WMAP9/2013\tablenotemark{b}                 & $70.0\pm 2.2$ &            &            &\checkmark & \checkmark & \checkmark &            &            &              \\
\WMAP9/2014\tablenotemark{c}                 & $69.4\pm 2.2$ &            &            &           &            &            & \checkmark & \checkmark & \checkmark   \\
\planck+\wmap Pol/2013\tablenotemark{d}      & $67.3\pm 1.2$ & \checkmark & \checkmark &           &            &            &            &            &              \\
ACT+SPT/2012\tablenotemark{e}                & --            &            &            &\checkmark & \checkmark & \checkmark &            &            &              \\
ACT+SPT/2013\tablenotemark{e}                & --            & \checkmark & \checkmark &           &            &            &            &            &    \\
ACT+SPT/2014\tablenotemark{e}                & --            &            &            &           &            &            &            & \checkmark & \checkmark   \\
\hline 
BAO/2012\tablenotemark{f}                    & --            &            &            &\checkmark & \checkmark & \checkmark &            &            &              \\
BAO/2013\tablenotemark{f}                    & --            & \checkmark & \checkmark &           &            &            &            &            &    \\
BAO/2014\tablenotemark{f}                    & --            &            &            &           &            &            & \checkmark & \checkmark & \checkmark  \\
\hline 
$H_0$, \citet{riess/etal:2011}               & $73.8\pm 2.4$ &            &            &           & \checkmark &            &            &            &              \\
$H_0$, \citet{riess:2014}                    & $73.0\pm 2.4$ &            & \checkmark &           &            & \checkmark & \checkmark &            & \checkmark   \\
\hline 
Combined $H_0$   (\kmsmpc)                   &               & 67.8       & 68.3       & 68.8      & 69.3       & 69.2       & 69.3       & 69.3       & \HubbleMean      \\
Combined $\sigma$   (\kmsmpc)                &               &  0.8       &  0.7       &  0.8      &  0.8       &  0.8       &  0.8       &  0.7       & \HubbleSigma         \\
\enddata
\tablenotetext{a}{Column H represents our current best-estimate concordance value of $H_0$, \Hubble, assuming flat $\Lambda$CDM.}
\tablenotetext{b}{\WMAP9/2013: From \citet{hinshaw/etal:2013}.  Assumes flat $\Lambda$CDM, with $\sum m_\nu=0$, $Y_{\rm He}=0.24$, $N_{\rm eff} = 3.04$ (instead of 3.046), and the January 2012 version of CAMB.}
\tablenotetext{c}{\WMAP9/2014: This paper. Assumes flat $\Lambda$CDM, with $\sum m_\nu=0.06$ eV, $Y_{\rm He}$ co-varies with $\Omega_b h^2$ per BBN, $N_{\rm eff} = 3.046$, and the March 2013 version of CAMB.}
\tablenotetext{d}{\Planck: From \citet{planck/16:2013}. Assumes flat $\Lambda$CDM. Likelihood ($2 \le \ell \le 2500$)+lensing+low-$l$ pixel-space polarization from the \WMAP9 likelihood.}
\tablenotetext{e}{ACT+SPT/2012: Uses ACT and SPT data that were available as of the final \WMAP9 release. ACT+SPT/2013: Uses data that were available as of the initial \planck\ release. ACT+SPT/2014: Uses data currently available for use in this paper.}
\tablenotetext{f}{BAO/2012: From Table 1 of \citet{hinshaw/etal:2013}. BAO/2013: From \S5.2 of \citet{planck/16:2013}. BAO/2014: From Table~\ref{tab:bao_data_tab} of this paper.}
\end{deluxetable*}
\capstarttrue

The weighted average results are reported in Table~\ref{tab:combination}.  The format of the table is as follows: the data sets we employ are listed in the first column; where that data set sufficiently constrains $H_0$ by itself, we quote that value in the second column.  Subsequent columns, labeled A through H, indicate different combinations of data, and their weighted average is given in the bottom row of each column.  Columns A and B contain \planck-based data sets \citep{planck/16:2013}, both without (Column A) and re-weighted with (Column B) the local $H_0$ value included.   Column C gives the \WMAP9+ACT+SPT+BAO result reported by \citet{hinshaw/etal:2013}, with no $H_0$ prior. Column D adds an $H_0$ prior, as reported by \citet{bennett/etal:2013}, and Column E gives the corresponding result re-weighted with the revised $H_0$ value in \citet{riess:2014}. Column F gives our new result based on (\WMAP9+BAO+$H_0$)/2014.  
Column G is the same as Column F, with the addition of ACT and SPT data, and the removal of the local $H_0$ prior.
Column H is the same as Column F, with the addition of ACT and SPT data.  As discussed below, Column H represents our current best-estimate concordance value of $H_0$, \Hubble.

We test consistency of the data in Figure~\ref{fig:wmap_planck_lcdm}.  To compare $H_0$, BAO, and CMB measurements, we present constraints in the space \{$H_0$, \dvrfidrd\} since this allows us to place $H_0$ measurements on the horizontal axis and BAO measurements orthogonally on the vertical axis.  Since the $z=0.57$ BAO measurements have the smallest errors (representing a 1\% measurement of the distance scale), and since the other BAO measurements are largely consistent with the $z=0.57$ measurement, we confine our plotted consistency checks to this BAO redshift only.  We add CMB data by projecting the Monte Carlo Markov Chain points into this space. 

Figure~\ref{fig:wmap_planck_lcdm} contains three panels with CMB constraints derived assuming \lcdm.  The top panel shows \wmap9/2014 constraints, the middle panel shows (\WMAP9+ACT+SPT)/2014, and the bottom panel shows (\Planck+ACT+SPT)/2013, on the same scale for ease of comparison.  To quantitatively assess consistency, we fit a 10-component two-dimensional (2D) Gaussian mixture function to the CMB chain points.  This function is evaluated numerically on a large 2D grid and contours that enclose 68.27\% and 95.45\% of the probability are computed and plotted.   (The accuracy of the Gaussian mixture can be checked by eye in Figures~\ref{fig:wmap_planck_lcdm}, \ref{fig:bao_h0_cmb1}, and \ref{fig:bao_h0_cmb2}, where we plot the
fitted contours over the chain points).  The Gaussian mixture provides an analytic 2D likelihood which we can combine with the 2D BAO+$H_0$ Gaussian likelihood (see below) to evaluate the best 2D fit.  

We note that neither the BAO nor the local Hubble constant measurements have extra variables in their likelihoods that are correlated with CMB measurements in other dimensions.  For example, neither the BAO measurement \dvrfidrd\ nor the local $H_0$ measurement depend on $\Omega_b h^2$, the scalar spectral index $n_s$, etc.  While it is true that $r_d = r_s(z_d)$ does depend on $\Omega_b h^2$, the quantity measured by the BAO, \dvrfidrd, is independent of $r_d$.  Given this lack of correlated hidden variables, we can determine the consistency of these likelihoods directly in this 2D space.

Since the BAO and $H_0$ data each constrain one dimension of our parameter space, we can combine them into a single likelihood to get a 2D Gaussian likelihood that constrains both \dvrfidrd\ and $H_0$.   We take $H_0 = 73.0\pm 2.4$ \kmsmpc\ \citep{riess:2014} and $\dvrfidrd = 2056\pm 20$ Mpc at $z=0.57$ \citep{anderson/etal:2013} and plot the combined likelihood as grey ellipses in Figures~\ref{fig:wmap_planck_lcdm}, \ref{fig:bao_h0_cmb1}, and \ref{fig:bao_h0_cmb2}.  We further note that the CMB likelihood for flat $\Lambda$CDM is well approximated by a single Gaussian in this space, producing contours that are virtually indistinguishable in Figure~\ref{fig:wmap_planck_lcdm}.   Under the assumption that both experiments are measuring the same cosmological quantities with different noise, the covariance matrix of the difference between the two measurements is the sum of their individual covariance matrices. 

For the \wmap9/2014 data in the top panel of Figure~\ref{fig:wmap_planck_lcdm}, we find the mean value of $H_0$ for the CMB-only likelihood to be 69.4 \kmsmpc.  The \wmap/2014 $H_0$ and $D_V$ values are entirely consistent with the BAO+$H_0$ likelihood; the difference has a $\chi^2$ of 2.5 for 2 degrees of freedom, with a probability to exceed (PTE) of 28\%.  Combining the three data sets gives a weighted mean of $H_0=69.5$ \kmsmpc, which is also clearly consistent with the individual data sets.  The maxima of the separate CMB and $H_0+$BAO likelihoods in this space are indicated with black filled circles, and the best fit to both distributions is indicated with an open square.  The best fit $H_0$ value for the CMB, local $H_0$, and BAO$_{z=0.57}$ likelihoods is also given in the figure.  Because the more sophisticated Gaussian mixture fit changes the results only very slightly, we consider the single Gaussian approximation to be acceptable for checking consistency of the fit.

Using an identical analysis for the (\wmap9+ACT+SPT)/2014 data, we find the mean of the CMB-only likelihood to be 71.1 \kmsmpc, with a $\chi^2$ for the difference of 3.8, with a PTE of 15\%. For the combined CMB+$H_0$+BAO$_{z=0.57}$ data, we find $H_0 = 69.8$ \kmsmpc.  Again, this is quite close to the Gaussian mixture result in the middle panel of Figure~\ref{fig:wmap_planck_lcdm}.

For the \planck-based CMB data in the bottom panel of Figure~\ref{fig:wmap_planck_lcdm},  we find the mean value of $H_0$ for the CMB-only likelihood to be 67.3 \kmsmpc.  This is only slightly less consistent with the BAO+$H_0$ likelihood;  the difference has a $\chi^2$ of 4.6 for 2 degrees of freedom and a PTE of 10\%.  The weighted mean of CMB+$H_0$+BAO$_{z=0.57}$ is 68.6 \kmsmpc.  This is 1.8$\sigma$ from the central value of the local Hubble prior,  and so is not significantly different.

As a further consistency check of the results in Figure~\ref{fig:wmap_planck_lcdm},  we list the deviations of the best fit point from each of the three likelihoods used (CMB, BAO$_{z=0.57}$, and $H_0$).  The deviation is given in number of standard deviations, which is directly computable for the $H_0$ and BAO data.  For the CMB, we compute the probability of the CMB Gaussian mixture likelihood to exceed the likelihood at the best fit point, and convert that into a number of standard deviations.  None of these values are above $2\sigma$, so the combined parameters are consistent with each individual likelihood.

\begin{figure} \begin{center}
\begin{tabular}{c}
\includegraphics[height=2.5in]{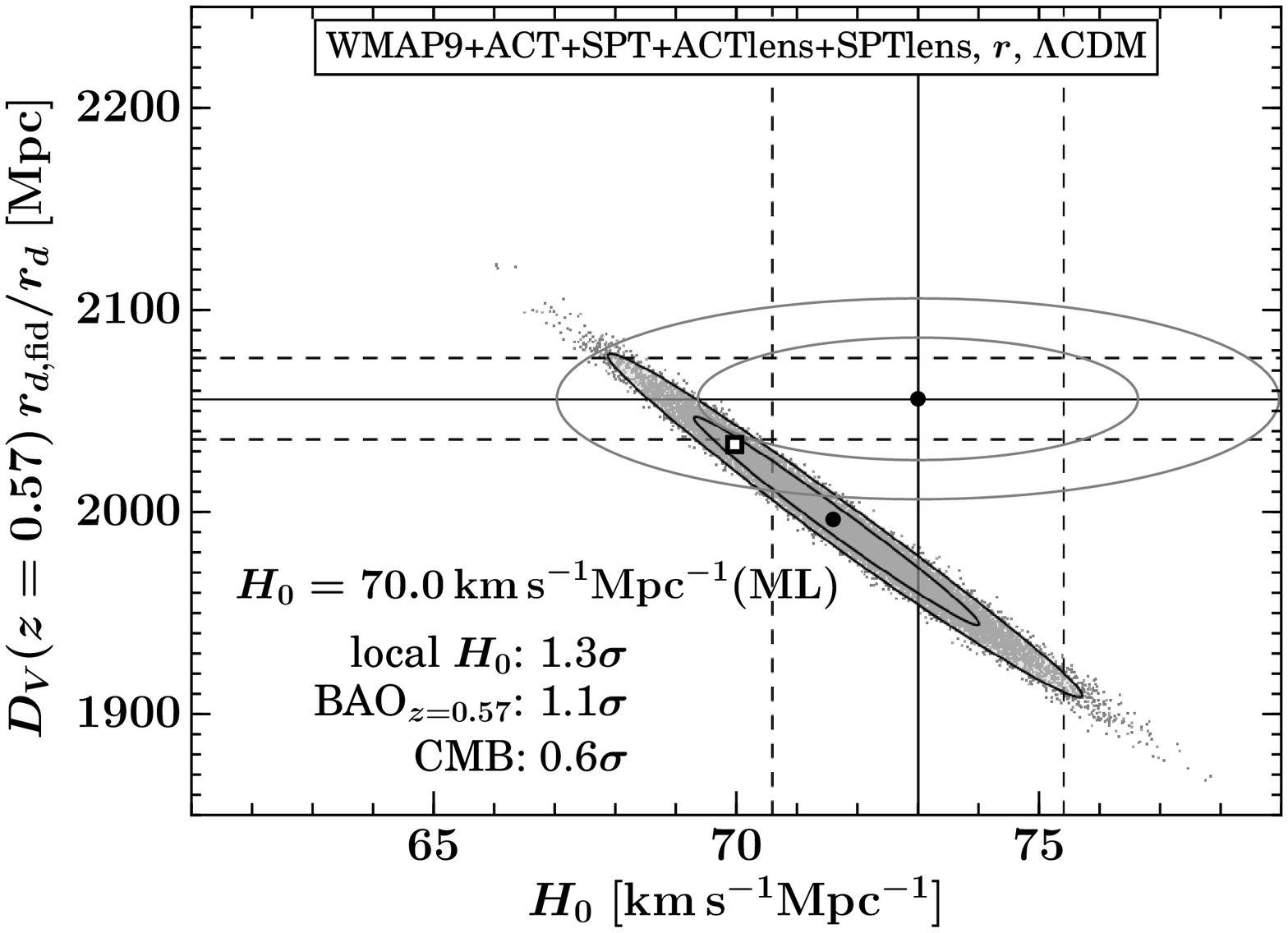} \\[-0.1in]
\includegraphics[height=2.5in]{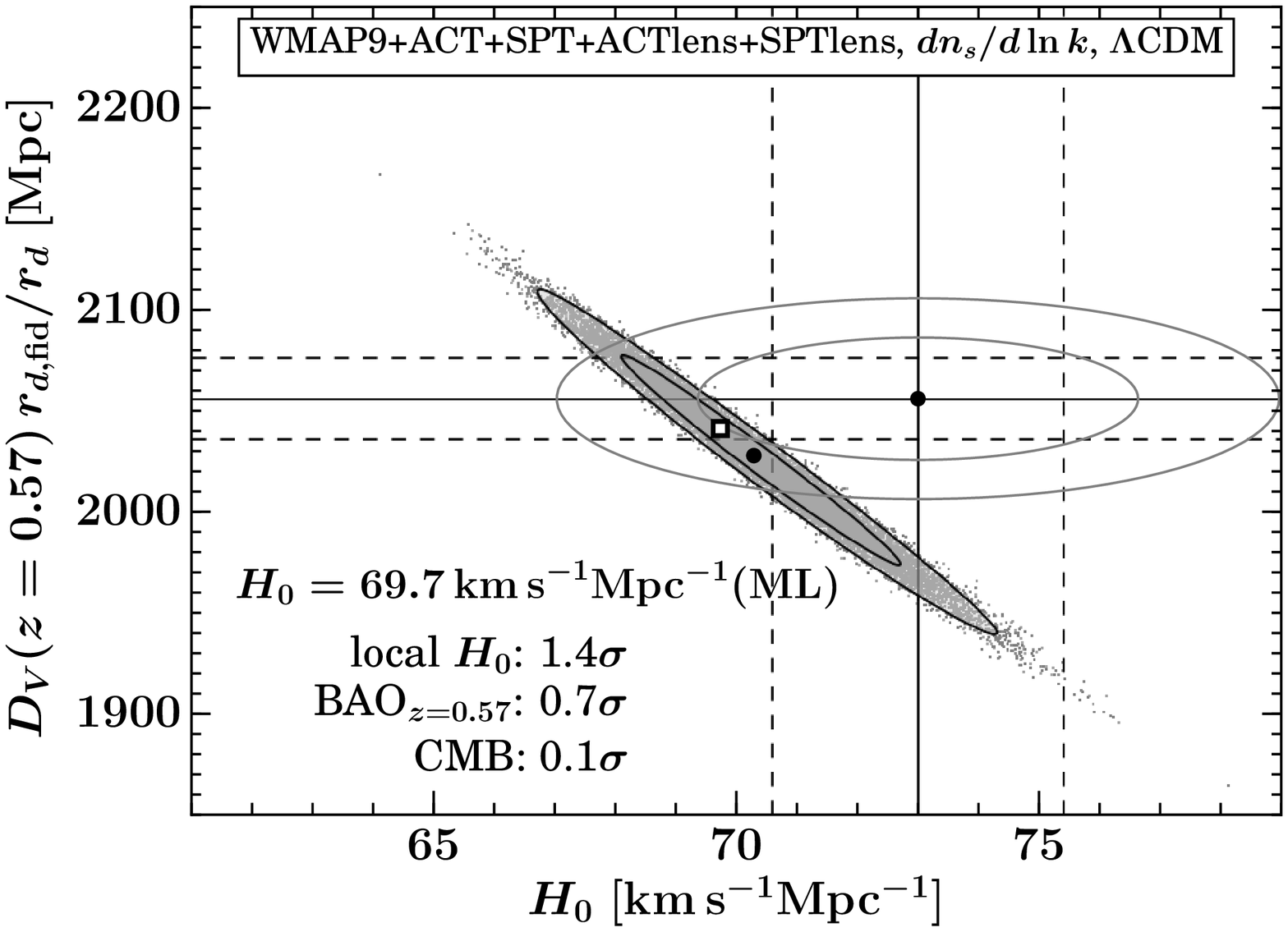} \\[-0.1in]
\includegraphics[height=2.5in]{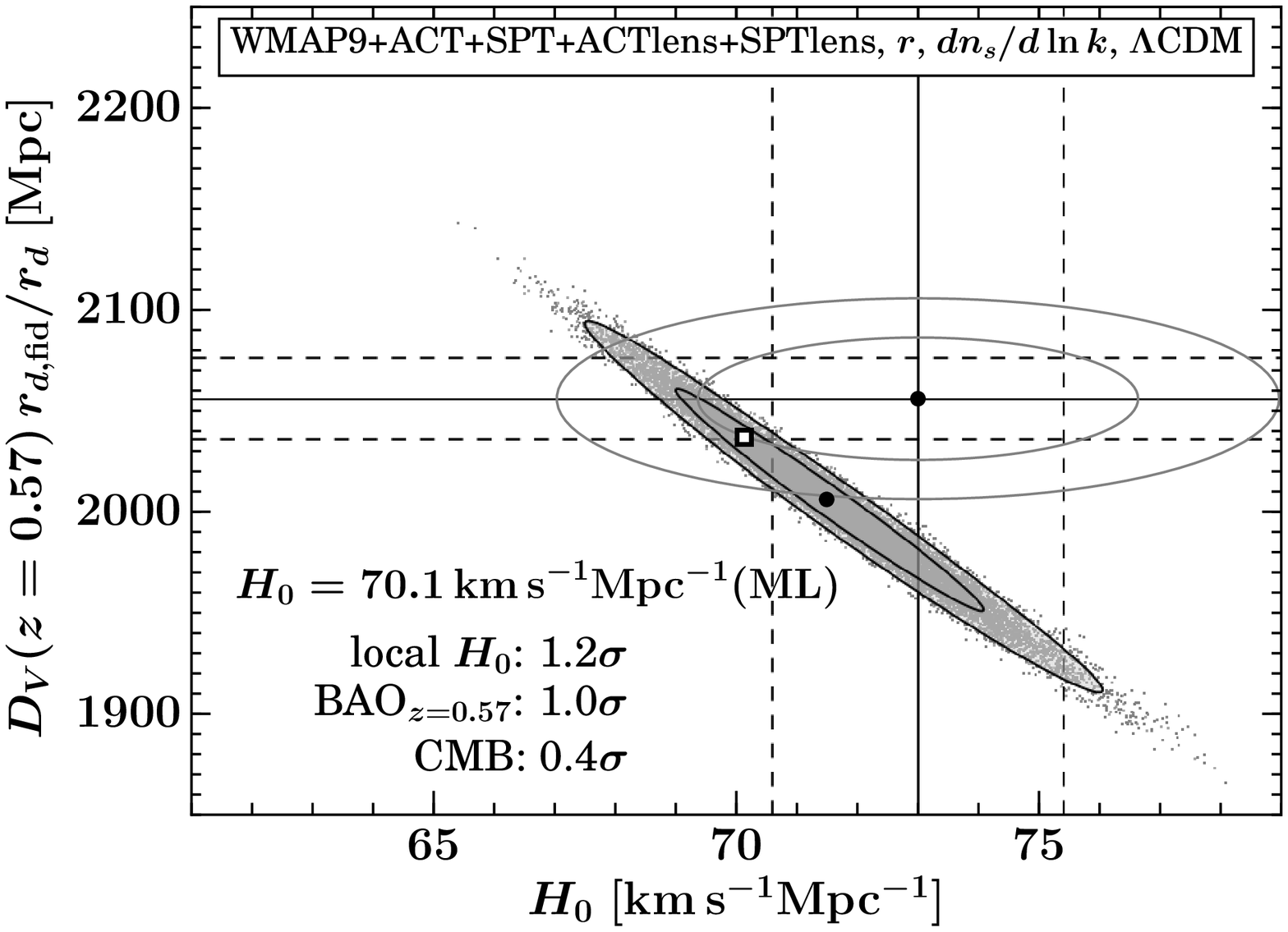} \\[-0.2in]
\end{tabular}
\end{center} 
\caption{Combination of $z=0.57$ BAO, local $H_0$, and CMB data.  Gray points and contours are from 
(\wmap9+ACT+SPT)/2014, as described in the text.
The black filled circle is the peak of the 2D marginalized CMB likelihood.  The concordance value, the open square, is maximum likelihood of the product of the 2D CMB, BAO, and $H_0$ likelihoods. 
\label{fig:bao_h0_cmb1}
}
\end{figure}

\addtocounter{figure}{-1}

\begin{figure}
\begin{center}
\begin{tabular}{c}
\includegraphics[height=2.5in]{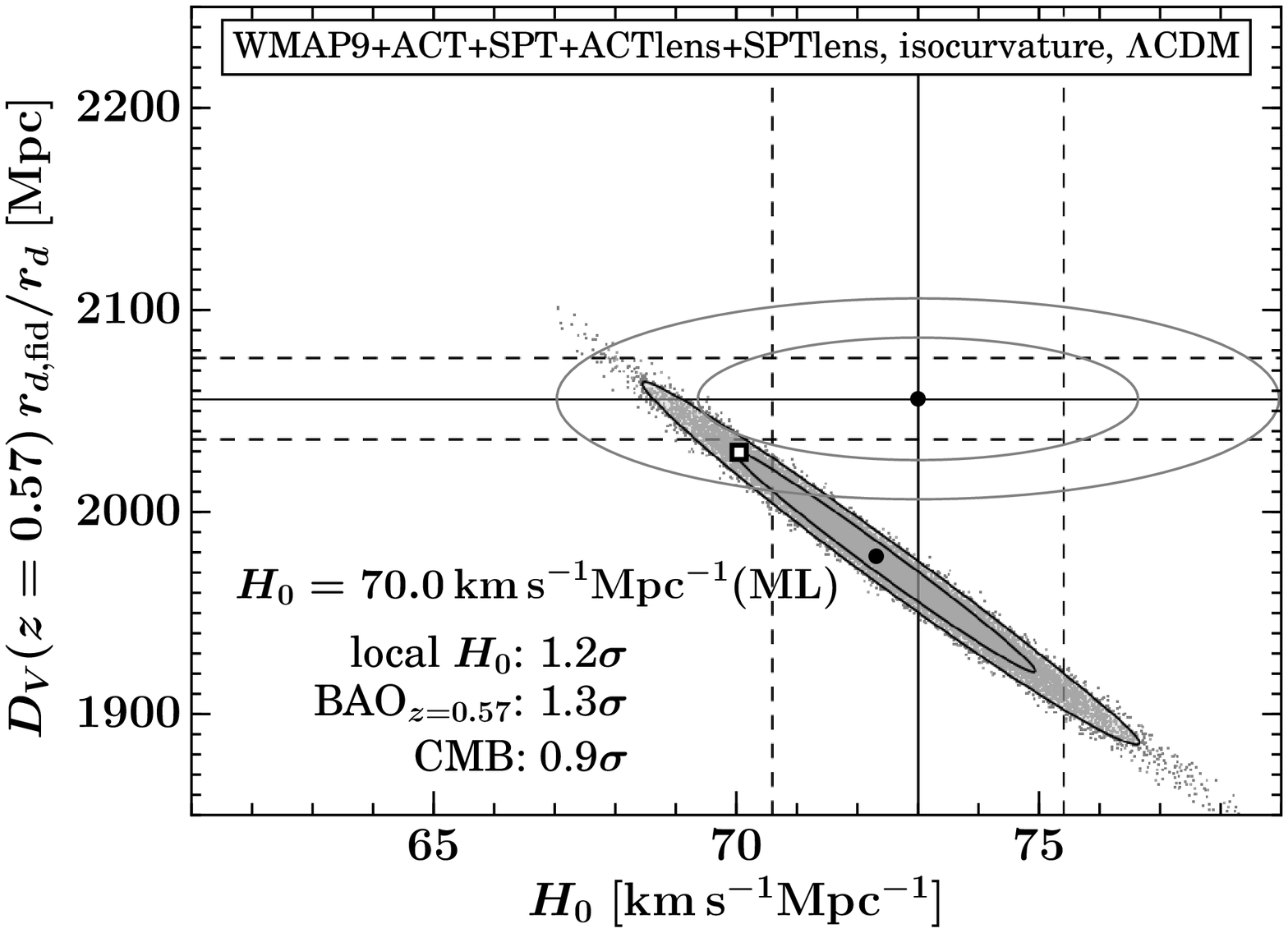} \\[-0.1in]
\includegraphics[height=2.5in]{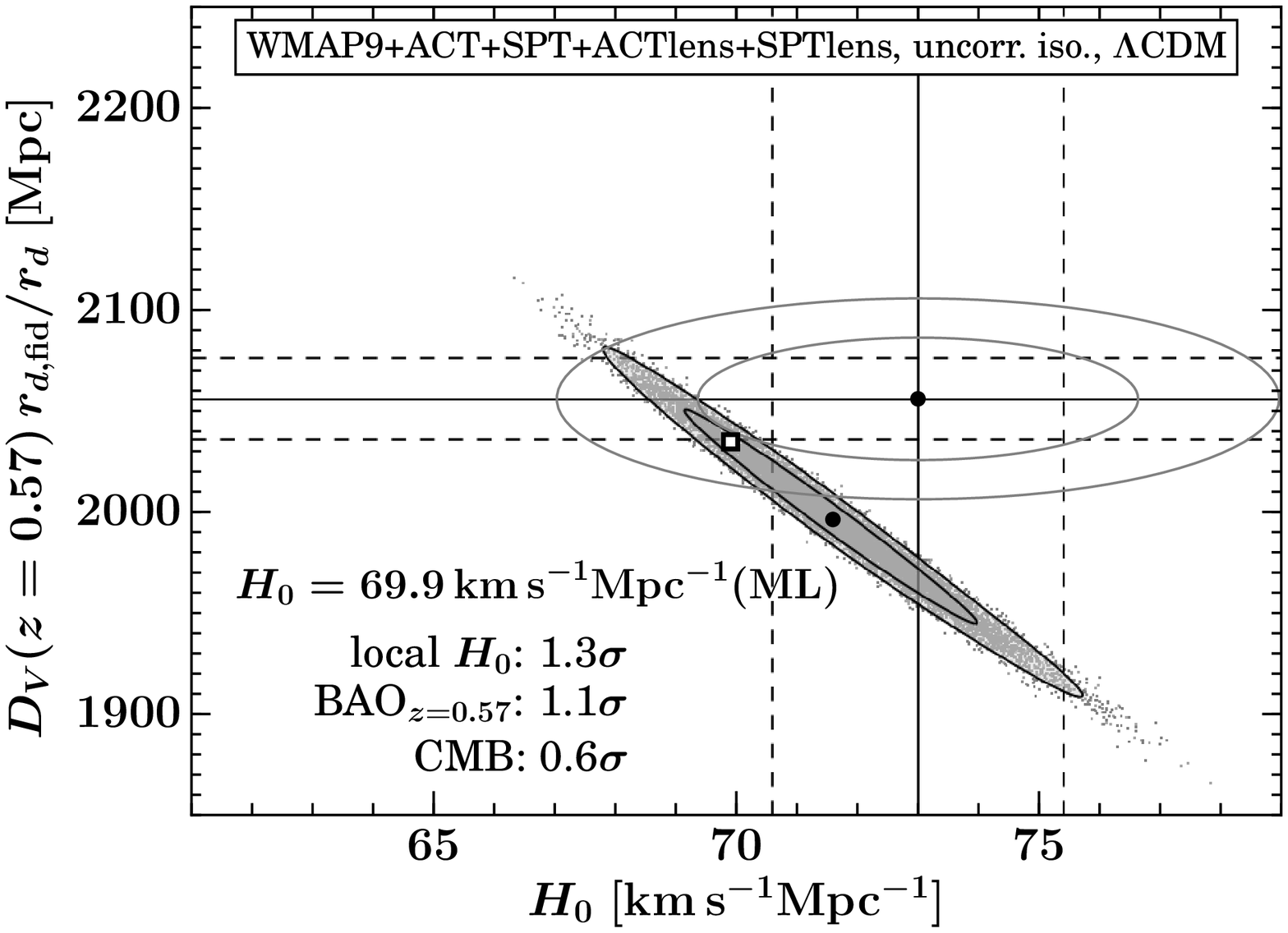} \\[-0.2in]
\end{tabular}
\end{center}
\caption{(Continued)}
\end{figure}

As shown in Figure~\ref{fig:wmap_planck_lcdm}, both \Planck\ and \WMAP\
data fall along the same degeneracy line. This is primarily because they provide
consistent measurements of the angular scale of the CMB peaks (see
\citet{page/etal:2003c,percival/etal:2002} for a discussion of these angular scales).
The \planck\ result itself prefers a slightly lower $H_0$ values than \wmap, but when combined 
with BAO data we see that the BAO tends to pull \planck\ upward and \wmap\ downward, 
yielding closer agreement between combined $H_0$ values for CMB+BAO+$H_0$ results. 
Therefore, it does not matter greatly whether we use the \wmap9+ACT+SPT or
\Planck+ACT+SPT data combination since they both constrain the ultimate $H_0$ fit value 
by virtue of following the same CMB degeneracy line.
Since the \planck\ team has
only had one data release so far and has not yet released revised chains reflecting
the correction for the 4~K cooler line effect mentioned in \S\ref{sec:planck},
we consider \wmap9+ACT+SPT CMB data in the remainder of this analysis.
We look forward to future releases of
\planck\ data, and in particular the updated chains and accompanying likelihood which 
will permit further independent analyses. 

In Figures~\ref{fig:bao_h0_cmb1} and \ref{fig:bao_h0_cmb2} we plot constraints
on $H_0$ and a BAO-measured distance to $z=0.57$ for various cosmological
models, to investigate the dependence on model.  We find that the constraint
from the CMB on all the models in Figure~\ref{fig:bao_h0_cmb1} is largely the
same; all these models predict the same expansion history of the universe.
This includes a tensor to scalar ratio $r$;
running of the spectral index with log wave number $k$ in Mpc$^{-1}$ given by $d n_s/d \ln k$; 
both running and $r$; 
and correlated and uncorrelated isocurvature modes.

\begin{figure} \begin{center}
\begin{tabular}{c}
\includegraphics[height=2.5in]{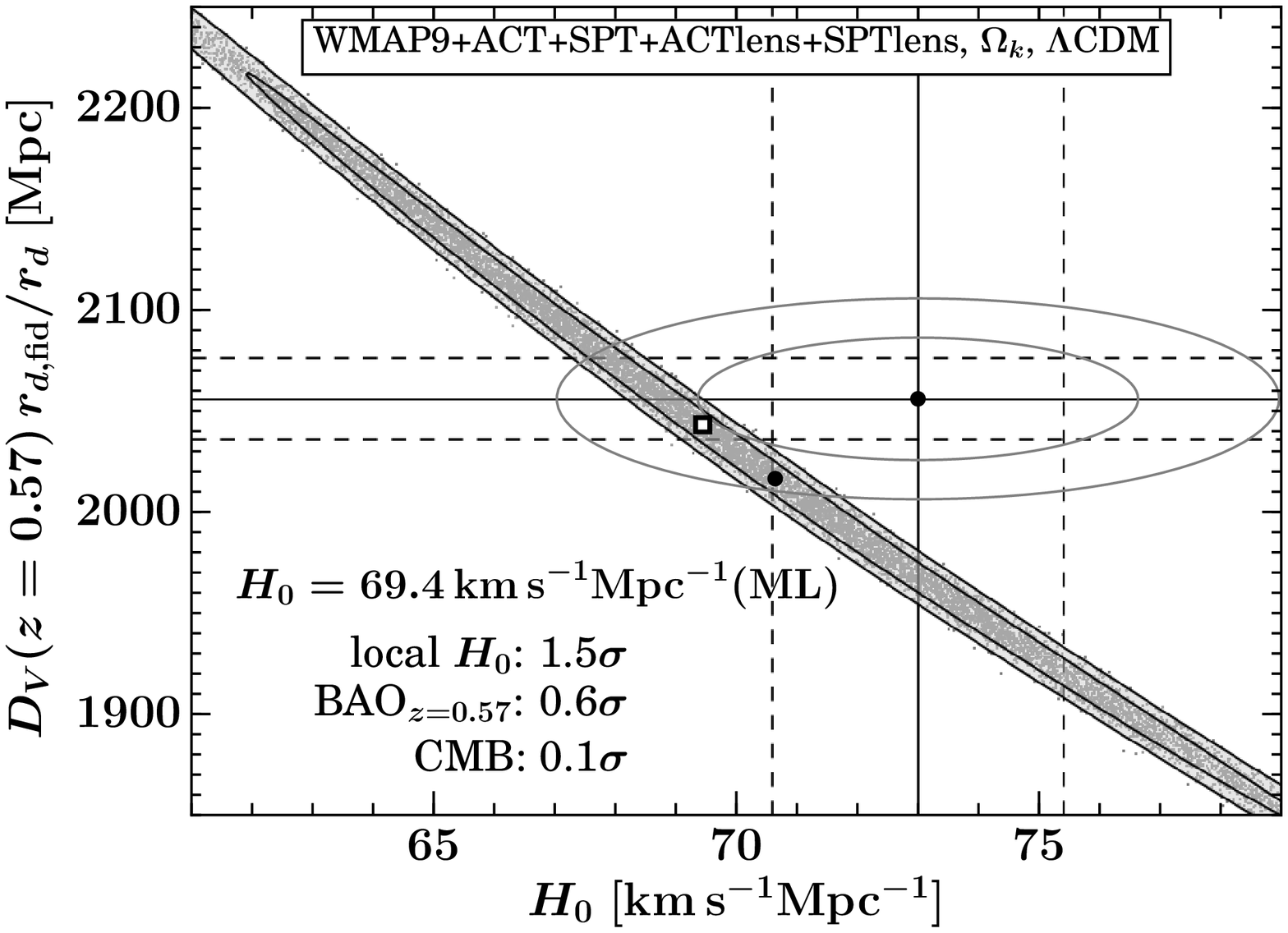} \\[-0.1in]
\includegraphics[height=2.5in]{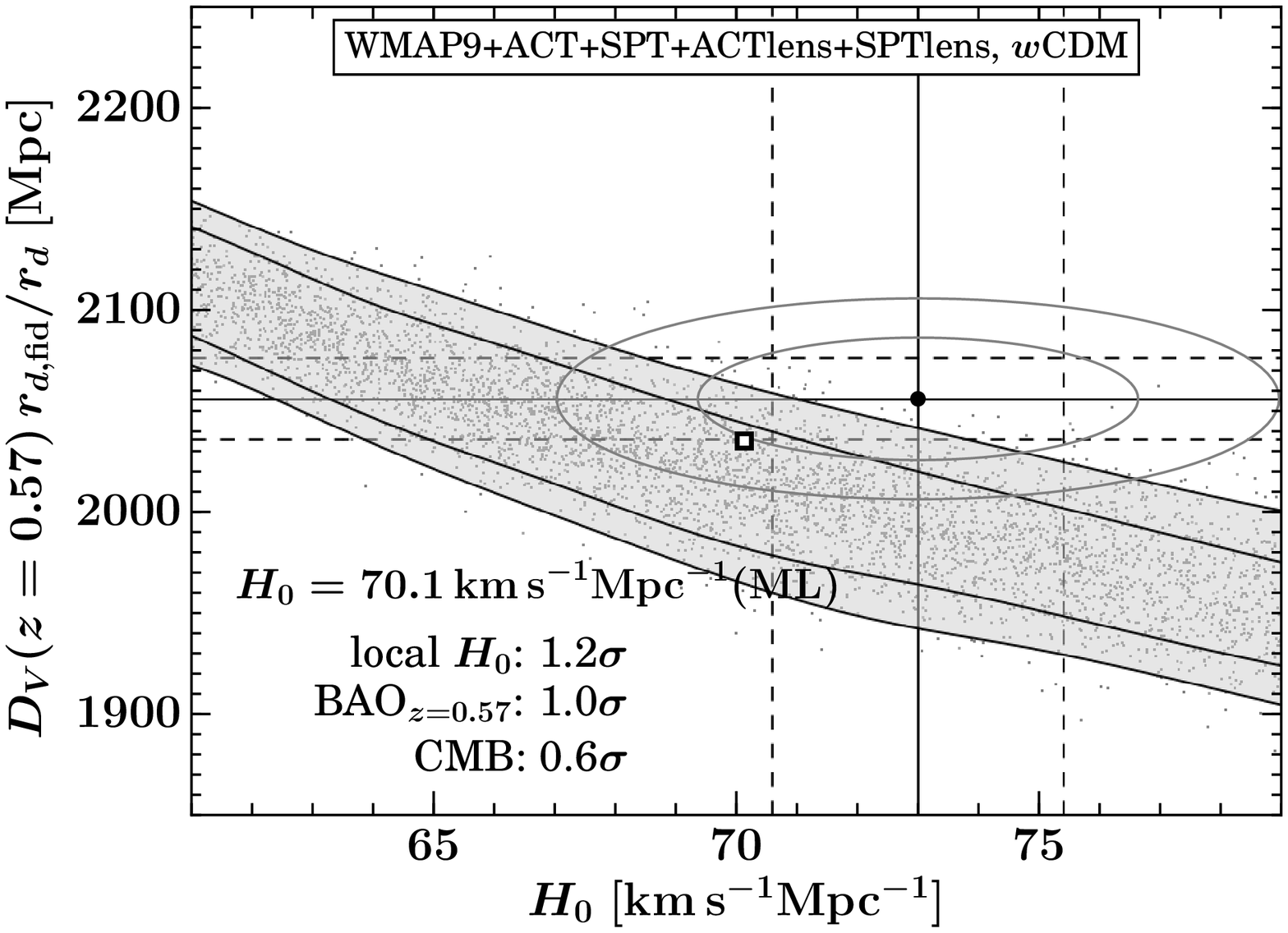} \\[-0.1in]
\includegraphics[height=2.5in]{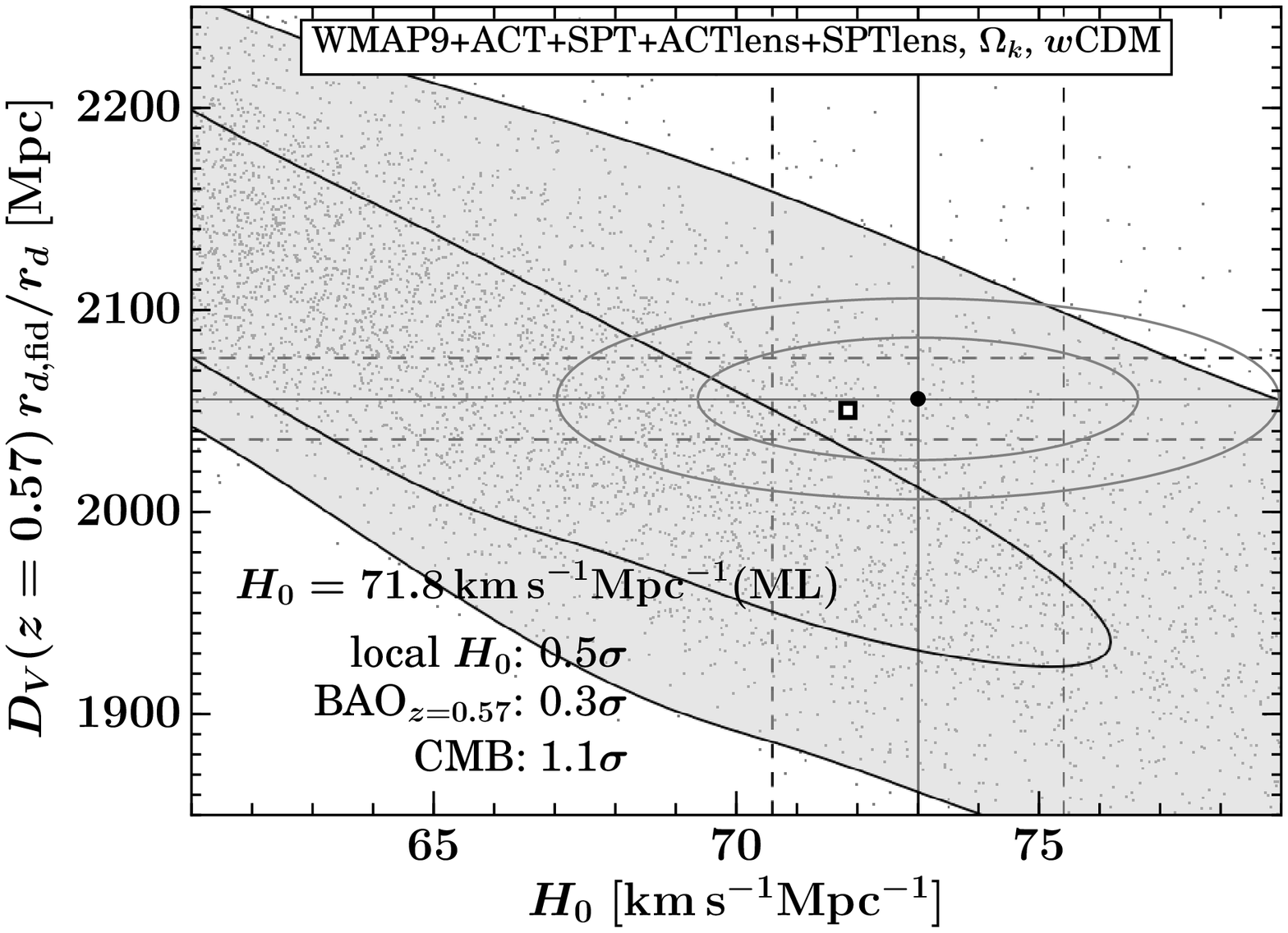} \\[-0.2in]
\end{tabular}
\end{center}
\caption{Combination of $z=0.57$ BAO, $H_0$, and CMB data, as in Figure~\ref{fig:bao_h0_cmb1}, but now with additional cosmological parameters not required by the data such as space curvature and/or an equation of state parameter $w \ne -1$, and/or massive neutrinos. 
For models only weakly constrained by CMB data the 2D Gaussian mixture approach used is not ideal so the contours should be viewed with caution.
\label{fig:bao_h0_cmb2}
}
\end{figure}

\addtocounter{figure}{-1}

\begin{figure}[t!]
\begin{center}
\begin{tabular}{c}
\includegraphics[height=2.5in]{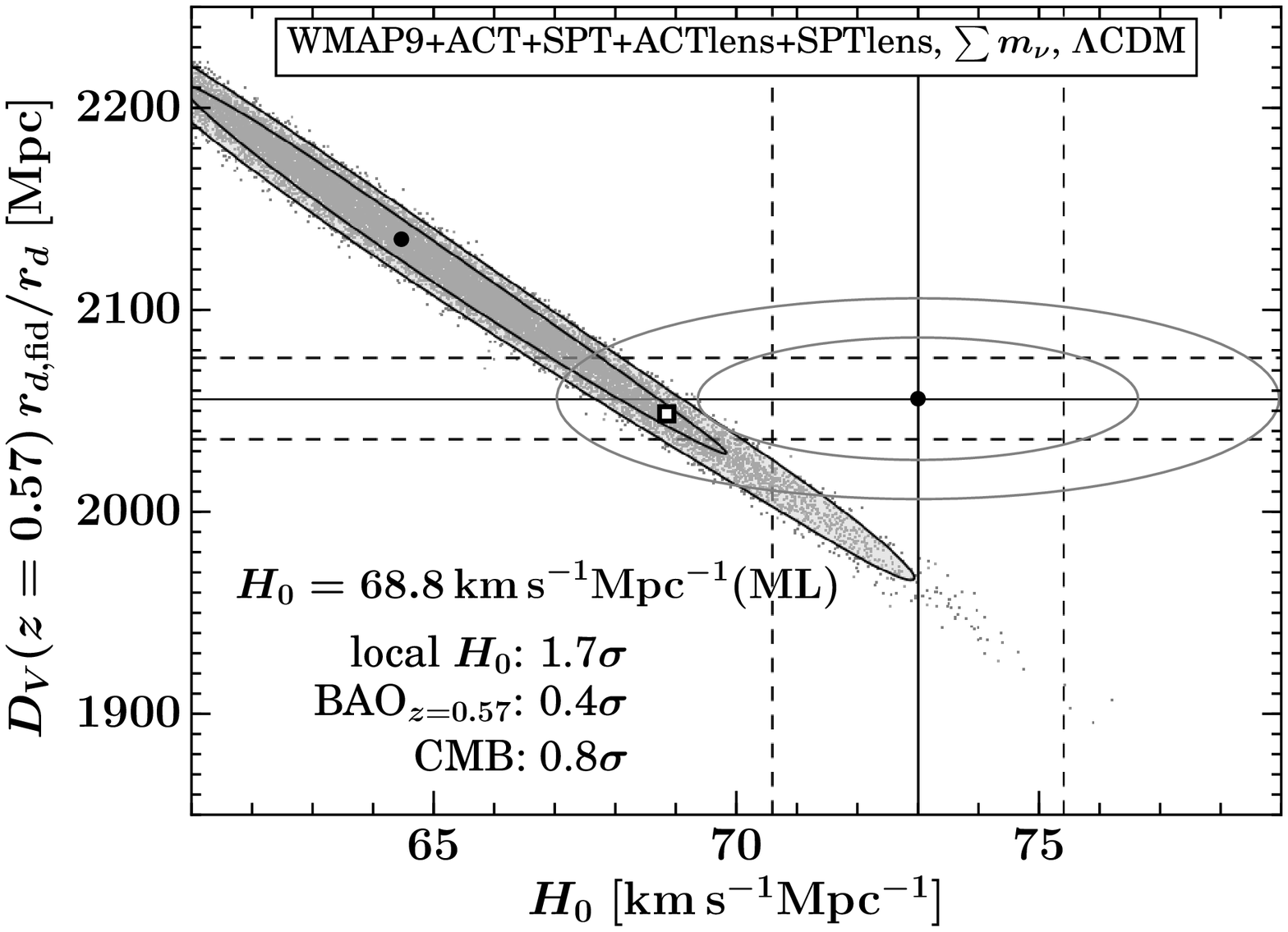} \\[-0.1in]
\includegraphics[height=2.5in]{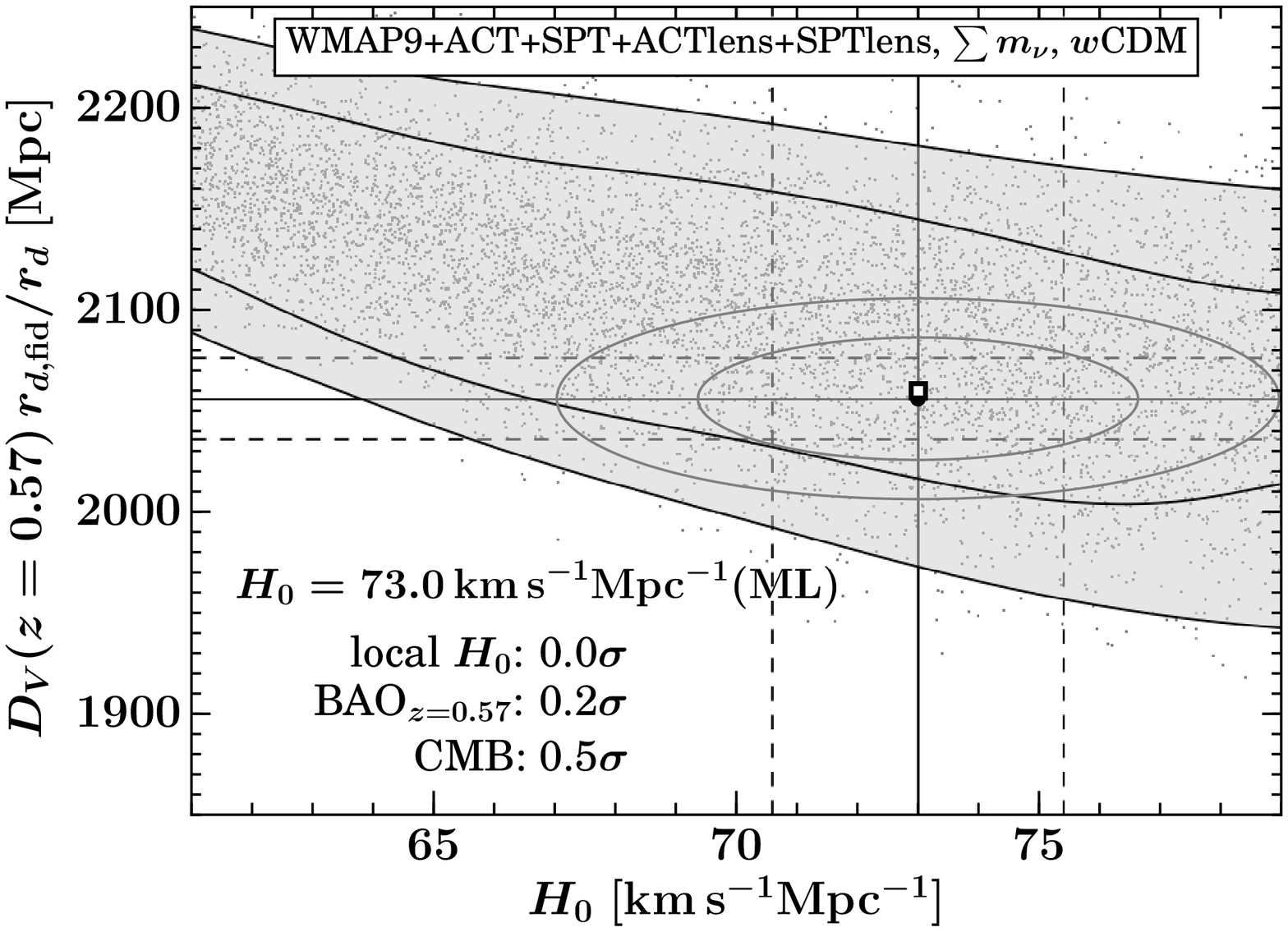} \\[-0.2in]
\end{tabular}
\end{center}
\caption{(Continued)}
\end{figure}

\addtocounter{figure}{-1}

\begin{figure}[t!]
\begin{center}
\begin{tabular}{c}
\includegraphics[height=2.5in]{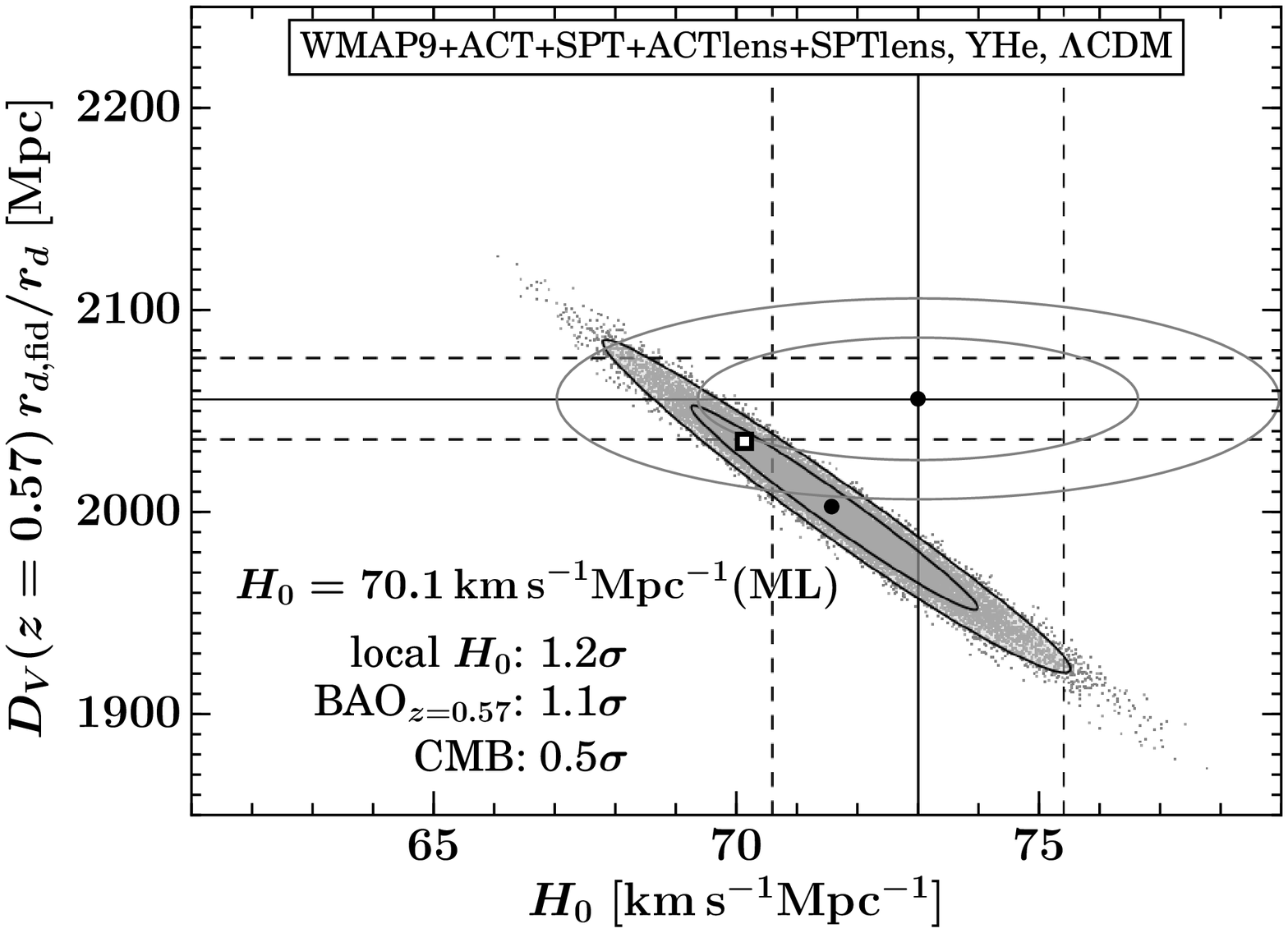} \\[-0.1in]
\includegraphics[height=2.5in]{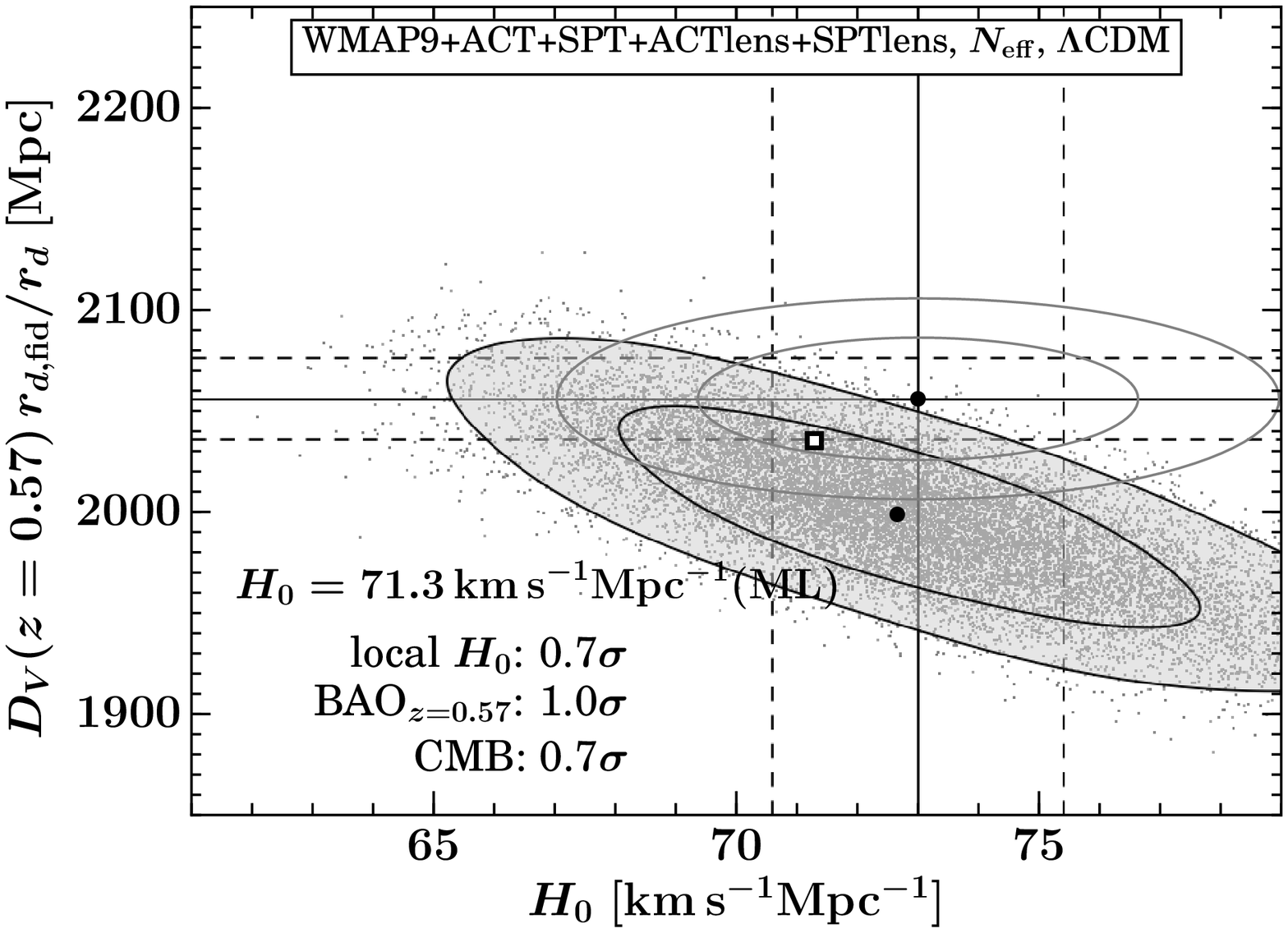} \\[-0.2in]
\end{tabular}
\end{center}
\caption{(Continued)}
\end{figure}

Figure~\ref{fig:bao_h0_cmb2} shows models with the following additional
parameters: 
curvature; 
equation of state $w\ne-1$;
curvature along with $w\ne-1$;
sum of the neutrino masses;
sum of the neutrino masses along with $w\ne-1$;
a varying primordial helium fraction $Y_{\rm He}$; 
and number of effective degrees of freedom in the early universe 
($N_{\rm eff}$).
While the models in Figure~\ref{fig:bao_h0_cmb1} all give substantially
similar results, the models that vary $N_{\rm eff}$ or the equation of
state in Figure~\ref{fig:bao_h0_cmb2} allow more freedom in the $D_V$, $H_0$ plane.
In some cases, this freedom provides very little CMB constraint at all, thus
allowing BAO+$H_0$ to dominate the combined result. In the case of the $w$CDM plot 
with massive neutrinos, all
that can really be concluded is that the CMB likelihood is relatively flat over
the BAO and $H_0$ best fit region, and still consistent with the \WMAP9
cosmological priors.

\begin{figure} 
\begin{center}
\includegraphics[width=0.5\textwidth]{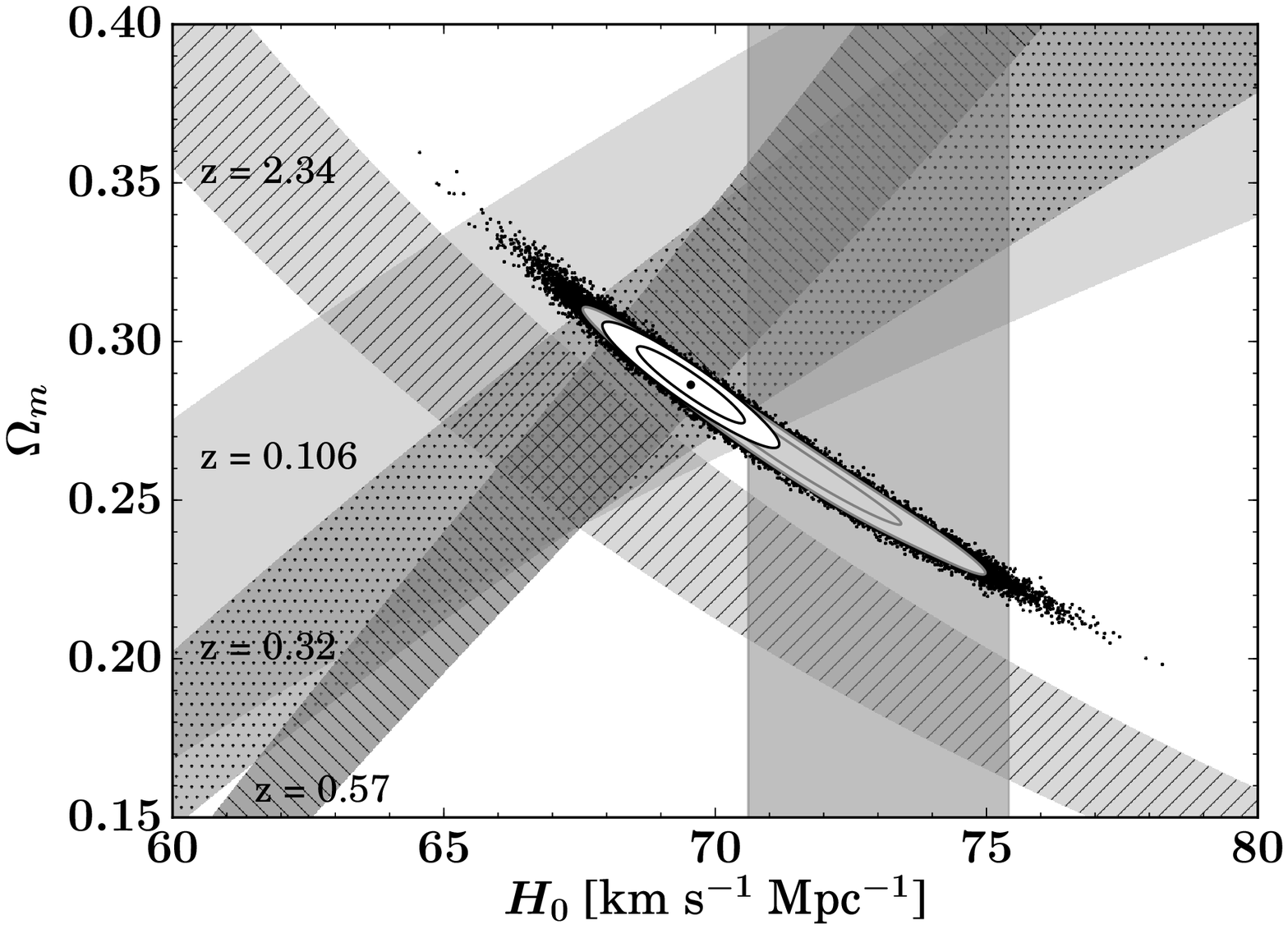}
\end{center} 
\caption{BAO $1\sigma$ constraints for redshifts 0.106
\citep[$r_s/D_V$ in][]{beutler/etal:2011}, 0.32 and 0.57 \citep[\dvrfidrd\ in][]{anderson/etal:2013}, and
2.34 \citep[$\alpha_\parallel^{0.7}\alpha_\perp^{0.3}$ in][]{delubac/etal:2014}. 
For these contours, the value of $r_d = r_s(z_d)$ is computed using
\citet{eisenstein/hu:1998} and corrected by a factor of 1.026 when comparing to
$r_d$ values computed using CAMB (everything except $z=0.106$).  The baryon
density is assumed to be $\Omega_b h^2 = 0.02223$ \citep{bennett/etal:2013};
changes of this value within the error bars do not substantially shift $r_d$ and
therefore the BAO contours.
The black points are from the (\WMAP9+ACT+SPT)/2014 $\Lambda$CDM Markov chain,
with 1 and 2$\sigma$ contours overlaid in gray.  The $H_0$
value with $1\sigma$ uncertainty from \citet{riess:2014} is indicated with 
the vertical shaded stripe.
For our previous figures, we only use the $z=0.57$ data set, since it is
consistent with the other BAO data and has the smallest error bars.  
The combination of all likelihoods shown is given by the white $1\sigma$ and
$2\sigma$ ellipses in the center of the figure.
Marginalizing over the other cosmological parameters, we find $\Omega_m = 0.286 \pm 0.008$ for all likelihoods shown. 
\label{fig:all_bao}
}
\end{figure}

In practice, \wmap\ and \planck\ chains that include BAO data (such as those presented in
Table~\ref{tab:combination}) utilize BAO measurements over a range of redshifts.
None of the currently released chains have included BAO data for $z>1$.
Representative BAO measurements at various redshifts (see Table ~\ref{tab:bao_data_tab})
are plotted in Figure~\ref{fig:all_bao}, for comparison purposes.  
Because these data mostly constrain the expansion history,
we plot these as constraints on $\Omega_m$ and $H_0$.  If we assume zero
curvature, then $\Omega_m + \Omega_\Lambda = 1$ (neglecting tiny amounts of
radiation and neutrinos), and so $\Omega_m$ and $H_0$ determine nearly the
entire expansion history of the universe, and certainly the region measured by
BAO.  To obtain the drag radius that is
directly measured by BAO, we assume the standard effective number of neutrinos
($N_{\rm eff} = 3.046$), and set $\Omega_b h^2 = 0.02223$ as measured by
\citet{bennett/etal:2013}.  Modification of these numbers slightly changes
$r_d$ and therefore the contours on this plot, but not substantially.
In this parameter space, it is clear that BAO measurements at different redshifts have different degeneracy 
directions, so that the inclusion of $z=2.34$ along with the lower redshifts
provides a fairly strong BAO-only constraint \citep[see also][]{addison/hinshaw/halpern:2013}.  
In addition, we plot the $1\sigma$ and $2\sigma$ likelihood contours for
the combination of BAO at all four redshifts, the local $H_0$ 
measurement, and (\WMAP9+ACT+SPT)/2014.
There is no significant tension between our mean value and these measurements.  
The $\chi^2$ of our best fit point with the four BAO likelihoods is 3.5 for four degrees of freedom.
The best fit $H_0 = $ \Hubble\ is $1.4\sigma$ away from the local $H_0$ measurement.
Finally, the best fit point is within the $1\sigma$ contour of the CMB points in this projection,
so it is also a good fit to the CMB data.

\section{Conclusions} \label{sec:conclusions}

1) Our combined dataset with CMB, BAO and distance ladder $H_0$ determinations results in
a best-estimate concordance Hubble constant value of \Hubble.
Given the currently available data, we argue that there is no compelling need for 
physics beyond standard flat $\Lambda$CDM to establish consistency between the $H_0$ values 
derived using three separate but complementary methods.

2) Following a similar analysis, the concordance value of $\Omega_m = 0.286 \pm 0.008$.

3) The previous $\approx 2.5 \sigma$ tension between the $H_0$ values determined by
\citet{planck/16:2013} and \citet{riess/etal:2011} has been recently mitigated somewhat (to $<2\sigma$) by
a decrease in the distance ladder value by roughly $0.3\sigma$ (\citet{riess:2014}, based on a recalibrated
megamaser distance) and a slight increase (also by  $\approx 0.3\sigma$) in the \Planck\ value derived 
assuming $\Lambda$CDM after correction for a 4~K cooler line instrumental effect 
\citep[][Appendix C.4]{planck/16:2013}.
Since the data incorporating the cooling line correction are not yet public, we
eagerly await the further insight into the Hubble constant
that will become available with the next \planck\ data release.

4) Within the 2D $D_V$ versus $H_0$ parameter space, CMB constraints tend to be confined along
a thin diagonal which is defined primarily by the acoustic scale of the CMB power spectrum.
As shown in Figure 1, both \Planck\ and \WMAP\ data provide very consistent measurements of the angular scale of the CMB peaks.
\planck\ data occupy a slightly different location along the degeneracy line than do \wmap\ data.
While the \planck\ result prefers a lower $H_0$ value than \wmap, the BAO
results pull \planck\ upward and \wmap\ downward, yielding CMB+BAO+$H_0$ results
that differ by roughly 1 \kmsmpc.
If the \Planck\ solution for $H_0$ in the next formal parameter release is significantly lower than the
\wmap9+ACT+SPT value, this would primarily indicate tension among the CMB data sets, 
which would require resolution before concluding that evidence for new physics is present.

5) We have shown that additional parameters are not necessary to have consistency
among the CMB, BAO, and $H_0$ results.  Should future measurements substantially
increase the tension, this could be resolved with additional model parameters.  
Figure~\ref{fig:bao_h0_cmb2} shows that this is most easily done with additional
relativistic degrees of freedom in the early universe ($N_{\rm eff}$), or an
equation of state not equal to $-1$, but we again emphasize that current data do
not require nonstandard values for these parameters.

6) It is vital to continue to improve the various measurements of $H_0$ in order to 
search for new physics. 
Of particular interest at this time would be a local $H_0$ 
determination with a precision smaller than that of our concordance value uncertainty.
Of further clear interest are decreased uncertainties in the high redshift (z$>2$) BAO
determinations.
We also recognize that there are always newer data
coming on-line (e.g., Sloan Digital Sky Survey with larger survey volume,  {\it Gaia} and WFC3 spatial 
scanning parallax results for better calibrator distances) and look forward to 
new contributions and constraints that these may provide.

We thank Adam Riess for useful comments. 
We also thank Daniel Eisenstein for suggesting the use of a Gaussian mixture model to 
fit the Markov chain points,
and an anonymous referee for suggesting helpful clarifications.
We acknowledge use of the  {\tt Scikit-learn}\footnote{
\href{http://scikit-learn.org/stable/index.html}
{http://scikit-learn.org/stable/index.html}} code \citep{pedregosa/etal:2011} for testing Gaussian mixture models.
This research was supported in part by NASA grant NNX14AF64G and by the Canadian Institute for Advanced Research (CIFAR). We acknowledge use of 
the HEALPix \citep{gorski/etal:2005} and CAMB \citep{lewis/challinor/lasenby:2000} packages.
This research has made use of NASA's Astrophysics Data System Bibliographic Services.
We acknowledge the use of the Legacy Archive for Microwave Background Data Analysis 
(LAMBDA). Support for LAMBDA is provided by NASA Headquarters.

\bibliographystyle{larson_apj}
\bibliography{wmap,planck,act,spt}

\end{document}